\documentclass[reprint,preprintnumbers,nofootinbib,superscriptaddress,amsmath,amssymb,aps,prd]{revtex4-2}
\pdfoutput=1
\usepackage{graphicx,hyperref,braket,orcidlink, algorithm2e, amssymb, amsmath, dsfont,subfigure}
\usepackage{lineno}
\usepackage{bm}
\usepackage{slashed}
\graphicspath{{figs/}}
\hypersetup{
	bookmarks=true,
	unicode=true,
	pdftoolbar=true,
	pdfmenubar=true,
	pdffitwindow=false,
	pdfstartview={FitH},
	pdftitle={QOC},
	pdfauthor={J.Y.~Araz, M.~Grau, F.~Ringer}, 
	pdfsubject={QOC},
	pdfcreator={J.Y.~Araz, M.~Grau, F.~Ringer},
	pdfproducer={}, 
	pdfkeywords={},
	pdfnewwindow=true,
	colorlinks=true,
	linkcolor=blue,
	citecolor=olive,
	filecolor=magenta,
	urlcolor=cyan
}

\begin{document}

\title{State preparation of lattice field theories using quantum optimal control}

\author{Jack Y. Araz\orcidlink{0000-0001-8721-8042}}
\email{jackaraz@jlab.org}
\affiliation{Thomas Jefferson National Accelerator Facility, Newport News, VA 23606, USA}
\affiliation{Department of Physics, Old Dominion University, Norfolk, VA 23529, USA}

\author{Siddhanth Bhowmick\orcidlink{0009-0006-9977-0508}}
\email{siddhanth.219301455@muj.manipal.edu}
\affiliation{Department of Computer Science \& Engineering, Manipal University Jaipur, Jaipur, 303007, India}

\author{Matt Grau\orcidlink{0000-0002-2684-6923}}
\email{mgrau@odu.edu}
\affiliation{Department of Physics, Old Dominion University, Norfolk, VA 23529, USA}

\author{Thomas J. McEntire\orcidlink{0000-0001-6914-1061}}
\email{tjmcenti@buffalo.edu}
\affiliation{Department of Physics, The State University of New York at Buffalo, Buffalo, NY 14260, USA}

\author{Felix Ringer\orcidlink{0000-0002-5939-3510}}
\email{fmringer@jlab.org}
\affiliation{Thomas Jefferson National Accelerator Facility, Newport News, VA 23606, USA}
\affiliation{Department of Physics, Old Dominion University, Norfolk, VA 23529, USA}


\preprint{JLAB-THY-24-4102}

\begin{abstract}
We explore the application of quantum optimal control (QOC) techniques to state preparation of lattice field theories on quantum computers. As a first example, we focus on the Schwinger model, quantum electrodynamics in 1+1 dimensions. We demonstrate that QOC can significantly speed up the ground state preparation compared to gate-based methods, even for models with long-range interactions. Using classical simulations, we explore the dependence on the inter-qubit coupling strength and the device connectivity, and we study the optimization in the presence of noise. While our simulations indicate potential speedups, the results strongly depend on the device specifications. In addition, we perform exploratory studies on the preparation of thermal states. Our results motivate further studies of QOC techniques in the context of quantum simulations for fundamental physics.
\end{abstract}

\maketitle
\tableofcontents

\section{Introduction}

Some of the most challenging open problems in fundamental physics, such as real-time and non-equilibrium dynamics, are anticipated to be beyond the capabilities of classical computing. The rapid advancement of quantum computing and the development of innovative algorithms motivate the exploration of quantum simulations to address these questions. A natural framework for this exploration is the Hamiltonian approach to quantum field theories developed by Kogut and Susskind~\cite{Kogut:1974ag}, where the theory is discretized on a spatial lattice, but time remains a continuous variable. For example, quantum algorithms to study scattering processes in scalar field theory, which scale polynomially with simulation parameters, have been developed in foundational work by Jordan, Lee, and Preskill~\cite{Jordan:2011ci, Jordan:2017lea}. In recent years, significant progress has been made toward simulating non-Abelian gauge theories like quantum chromodynamics (QCD), which describes the strong force in nature. We refer the reader to Refs.~\cite{_imkovic_2017, Ercolessi:2017jbi, Klco:2018kyo, Chandrasekharan:1996ih, Hauke:2013jga, Banerjee:2012pg, Banuls:2019bmf, Shaw:2020udc, Barata:2020jtq, Cohen:2021imf, Bauer:2021gek, deJong:2021wsd, 
Angelides:2023bme, Honda:2021aum, Liu:2022grf, Barata:2023jgd, Magnifico:2019kyj, Li:2024nod, Farrell:2024fit, Barata:2022wim, Davoudi:2022uzo, Florio:2023dke, Gustafson:2023xpe, PhysRevA.109.062422, Florio:2024aix, Echevarria:2020wct, Chakraborty:2020uhf, Zohar:2015hwa, Ebner:2024mee, Belyansky:2023rgh,Nguyen:2021hyk, Thompson:2021eze, Bauer:2023qgm, Briceno:2023xcm, ARahman:2022tkr, Meth:2023wzd, Alexandru:2023qzd, Grieninger:2024cdl, Araz:2022tbd, PhysRevA.109.062422, Araz:2022zxk, Czajka:2021yll, Araz:2022haf, Carrillo:2024chu, Khor:2023xar, Araz:2023ngh, Farrell:2023fgd, Spagnoli:2024mib} for recent studies along these lines. One of the most challenging aspects of these algorithms is the preparation of initial states for the scattering process and, analogously, any correlation function relevant for describing non-perturbative aspects of dynamical processes in fundamental physics. Similar challenges apply to the preparation of thermal states, which are crucial for improving our understanding of systems such as the quark-gluon plasma that can be probed experimentally at high-energy collider experiments. Quantum simulations of relevant lattice field theories are currently limited by hardware noise. Gate errors and the decoherence of the quantum hardware constrain the complexity of quantum circuits that can be reliably executed to prepare states of lattice field theory and perform the subsequent time evolution and measurements.

To address these challenges, Variational Quantum Algorithms (VQA) have been proposed~\cite{Peruzzo2014, Bharti2022, Cerezo:2021aa, TILLY20221}, which allow for relatively shallow circuits. This approach employs a parameterized unitary quantum circuit to prepare the desired state by optimizing gate parameters with respect to the expectation value of the Hamiltonian in question. Despite the considerable success of these algorithms, the coherence time of near-term quantum devices remains a significant constraint. For instance, at the time of this study, the \texttt{ibm\_kyiv} quantum computer shows a median two-qubit gate duration of $561.778$ ns. For example, given that the typical coherence time of IBM's Eagle quantum processors is around $100$ $\mu$s, approximately $200$ to 250 two-qubit gates can be applied reliably\footnote{Gate time information for each quantum device has been presented in IBM Quantum Platform, under Systems.}. This poses a significant limitation for simulating lattice field theories. Furthermore, variational algorithms are prone to the problem of vanishing gradients, known as barren plateaus~\cite{McClean:2018jps, Ragone:2023qbn, Diaz:2023uuo, Fontana:2023wnj}. Additionally, local traps can hinder an efficient optimization~\cite{PhysRevLett.127.120502}.

Once a gate-level circuit instruction is submitted to a quantum device, it undergoes a series of compilation steps to convert gate-based instructions into machine language. During this process, each gate is converted to a series of predetermined pulse sequences designed to realize the gate and achieve a specific fidelity. In recent years,  Quantum Optimal Control (QOC) techniques have been proposed to mitigate this overhead by enabling a gate-free state preparation~\cite{Meitei:2021aa}. This approach operates directly at the level of hardware-specific pulses that are optimized to prepare an approximate unitary to prepare the desired state. This AI-driven workflow has been successfully applied to finding molecular ground states in several recent studies~\cite{Meitei:2021aa, Liang:2022gvt, PRXQuantum.2.010101, PhysRevApplied.19.064071, Sherbert:2024jdq, PhysRevResearch.5.033159, 9996174, 10.3389/frqst.2023.1273581, PhysRevX.7.021027, Long:2024nzk, Asthana:2022pgu, Asthana2023, Petruhanov:2023max,Entin:2024jmy}. QOC optimizes control parameters of the device Hamiltonian using a series of time-dependent pulses and evolving the Hamiltonian for a specific duration. The minimum time required to achieve the ground state with the desired accuracy is known as the minimum evolution time (MET)~\cite{PhysRevApplied.19.064071} or the quantum speed limit~\cite{Deffner_2017}. Once the MET is achieved or closely approximated, QOC landscapes are free from local minima, which can be a significant advantage compared to gate-based VQAs~\cite{russell2016quantum}.

In this study, we aim to demonstrate that QOC can significantly reduce the time required for the ground state preparation of lattice field theories on superconducting quantum devices. To the best of our knowledge, QOC techniques have not been explored in this context before. Specifically, we will focus on the Schwinger model~\cite{Schwinger1962, Coleman1975}, a U(1) gauge theory coupled to fermionic matter, corresponding to quantum electrodynamics in 1+1 dimensions. The Schwinger model is frequently studied because of its similarities with QCD, including confinement and chiral symmetry breaking. It can be expressed purely in terms of fermionic degrees of freedom with an asymmetric long-range interaction. Generally, the main challenge of QOC techniques is the scalability of optimizing pulse sequences for a large number of qubits. For confining lattice field theories like the Schwinger model, the correlation between fermions decays exponentially due to the screening of electric charges. This indicates that the pulse level optimization only needs to be carried out for a certain number of qubits. As demonstrated in recent work for the Schwinger model with more than 100 qubits~\cite{Farrell:2023fgd, Farrell:2024fit}, a truncation of the interaction between fermions at long distances allows for state preparation using scalable variational algorithms. While we leave the exploration of the design of scalable algorithms at the pulse level for future work, these results indicate that a scalable approach may also be achievable using QOC techniques while allowing for reduced coherence time requirements.
Additionally, we extend our studies in this work to the preparation of thermal states of the Schwinger model. Thermal states are described by a density matrix, making their preparation challenging using quantum computers. We employ an approach based on two variational ansätze with an intermediate measurement, which is optimized by minimizing the free energy. Using gate-level optimization, this algorithm was proposed in Ref.~\cite{Selisko:2022wlc}, which we extend in this work by using QOC techniques.

The remainder of this paper is organized as follows. In section~\ref{sec:schwinger}, we will review the Hamiltonian formulation of the Schwinger model, which is the exemplary lattice field theory considered throughout this work. In section~\ref{sec:QOC}, we will introduce the QOC ansatz for the state preparation based on the pulse-level control of a transmon Hamiltonian. In section~\ref{sec:state-prep-qoc}, we will discuss the optimization algorithm, present numerical results for the ground state preparation of the Schwinger model, and explore hardware constraints such as the dependence on device parameters and noise. We extend these calculations to prepare thermal states using a QOC ansatz in section~\ref{sec:thermal-state}. We conclude and present an outlook in section~\ref{sec:conclusions}.

\section{The Schwinger model}\label{sec:schwinger}

\begin{figure}[t]
    \centering
    \includegraphics[width=0.9\linewidth]{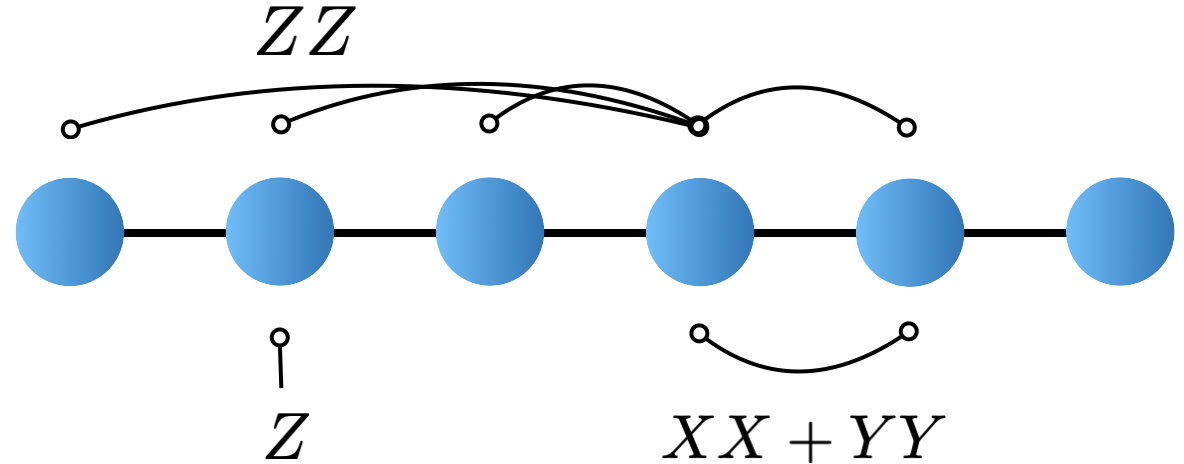}
    \caption{\it Schematic illustration of the onsite, nearest-neighbor, and long-range interactions of the Schwinger model expressed as a spin model for $N=6$ lattice sites.}
    \label{fig:SchwingerLattice}
\end{figure}

The Lagrangian of the massive Schwinger model, quantum electrodynamics in 1+1 dimensions, with a topological $\theta$-term is given by
\begin{equation}
    \mathcal{L}=\bar{\psi}(i \slashed{D}-m) \psi-\frac{1}{4} F^{\mu \nu} F_{\mu \nu} + \frac{e \theta}{4 \pi} \epsilon_{\mu \nu} F^{\mu \nu}\,.
\end{equation}
Here, $\psi$ denotes a two-component fermion field with mass $m$. The field strength tensor can be written in terms of the U(1) gauge field as $F_{\mu\nu}=\partial_\mu A_\nu-\partial_\nu A_\mu$, and the covariant derivative is given by $D_\mu=\partial_\mu-eA_\mu$, where $e$ is the electric charge. Moreover, $\theta$ is the topological angle and $\epsilon_{\mu\nu}$ is an asymmetric tensor. Using the axial gauge with $A_0=0$, staggered fermions, and a Jordan-Wigner transformation, the Schwinger model can be discretized on a spatial lattice~\cite{Kogut:1974ag}. In 1+1 dimensions, the U(1) gauge field is not a dynamical degree of freedom and is fully constrained by Gauss's law. Therefore, the gauge field can be replaced with an asymmetric long-range interaction such that the entire Hamiltonian can be written as a spin model~\cite{Hamer:1997dx, Martinez:2016yna}. Using open boundary conditions and a vanishing electric background field, we can write the Hamiltonian as
\begin{equation}\label{eq:HSchwinger}
    \hat H=\hat H_{ZZ}+\hat H_{\pm}+\hat H_Z \,.
\end{equation}
The first term contains the asymmetric all-to-all $ZZ$ interaction of the spins. It is given by
\begin{equation}\label{eq:HZZ}
    \hat H_{ZZ}=\frac{J}{2}\sum_{m=1}^{N-2} \sum_{n=m+1}^{N-1} (N-n) Z_m Z_n \,.
\end{equation}
Here, $N$ is the total number of lattice sites or qubits and $J=e^2a/2$, where $a$ is the lattice spacing. The second term in Eq.~(\ref{eq:HSchwinger}) is a hopping term between neighboring lattice sites
\begin{equation}
    \hat H_{\pm}=\sum_{n=1}^{N-1} \left(\frac{1}{2a} - \frac{(-1)^n m\sin\theta}{2}\right) (X_n X_{n+1}+Y_nY_{n+1})\,.
\end{equation}
Lastly, the third term of the Hamiltonian is given by
\begin{equation}
    \hat H_Z = \frac{m}{2}\cos\theta\sum_{n=1}^{N}(-1)^n Z_n-\frac{J}{2}\sum_{n=1}^{N-1} (n\;\textrm{mod}\; 2)\sum_{l=1}^n Z_l\,.
\end{equation}
We refer the reader to Refs.~\cite{Martinez:2016yna, Chakraborty:2020uhf} for more details.

\section{Quantum optimal control (QOC)~\label{sec:QOC}}

Throughout this work, we consider superconducting qubits, which can be modeled with a transmon Hamiltonian~\cite{PhysRevA.76.042319, PhysRevB.77.180502}. We split the Hamiltonian into a time-independent device and a time-dependent control Hamiltonian. The first part describes the physics of a set of coupled anharmonic oscillators and can be written as 
\begin{eqnarray}
    \hat{H}_D &=& \sum_i \omega_i\, \hat a^\dagger_i \hat a_i - \sum_i \frac{\delta_i}{2} \hat a^\dagger_i \hat a^\dagger_i \hat a_i \hat a_i \nonumber \\
    &+& \sum_{i,j} g_{ij} \left(\hat a^\dagger_i \hat a_j + {\rm h.c.}\right)\ .\label{eq:drive}
\end{eqnarray}
Here, $\omega_i$ represents the transition frequency between the $|0\rangle$ and $|1\rangle$ states of the $i^{\rm th}$ qubit, $\delta_{i}$ denotes its anharmonicity that leads to a different level spacing of the states $\ket{n\geq 2}$, and $g_{ij}$ is the constant coupling rate between the $i^{\rm th}$ and $j^{\rm th}$ qubit. The sum of the last term in Eq.~\eqref{eq:drive} runs over pairs of coupled qubits of a given device, which will be discussed in more detail below. The operators $\hat a^\dagger,\hat a$ are raising and lowering operators, respectively, which satisfy the commutation relation $[\hat a,\hat a^\dagger]=1$. Within the rotating wave approximation, the time-dependent control Hamiltonian is given by
\begin{eqnarray}
    \hat{H}_C(t) = \sum_i \Omega_i(t) \left( e^{i v_i t} \hat a_i + e^{-i v_i t} \hat a^\dagger_i \right)\ . \label{eq:control}
\end{eqnarray}
Here, $\Omega_i(t)$ is the time-dependent amplitude, and $v_i$ is the drive frequency, of the $i^{\rm th}$ qubit. Combining these equations yields the total optimal control Hamiltonian, $\hat{H}_{\rm OC}(\mathbf{\Omega}, \mathbf{v}, t)=\hat{H}_D+\hat{H}_C(t)$. Here, we denote the set of amplitudes and frequencies by $\mathbf{\Omega}, \mathbf{v}$, respectively. Note that Eq.~\eqref{eq:control} involves a sum over Hamiltonians acting on single qubits. Therefore, multi-qubit operations are achieved through the inter-qubit coupling term in Eq.~\eqref{eq:drive}. Depending on the device, the parameters in Eq.~\eqref{eq:drive} may also be tunable within device-dependent constraints. However, in this study, we will assume that the device Hamiltonian in Eq.~\eqref{eq:drive} is fixed. Any gate operation that can be realized with a quantum computer can be constructed by applying suitably chosen pulses defined in terms of the parameters $\mathbf{\Omega}, \mathbf{v}$. In the next section, we will instead directly prepare the ground state of the Schwinger model by adjusting the parameters of the control Hamiltonian, which enables a gate-free state preparation. 

The time evolution of the state is governed by the time-dependent Schr\"{o}dinger equation
\begin{eqnarray}
    \frac{\rm d}{{\rm d}t} |\Psi(t)\rangle = -i \hat{H}_{\rm OC}(\mathbf{\Omega}, \mathbf{v}, t) |\Psi(t)\rangle\ .\label{eq:schrodinger}
\end{eqnarray}
Its solution, which corresponds to the variational pulse-level ansatz employed in the next section, is given by
\begin{eqnarray}
    |\Psi(\mathbf{\Omega}, \mathbf{v})\rangle = \underbrace{\hat{\mathcal{T}} \exp\left[ -i\int_0^T{\rm d}t\, \hat{H}_{\rm OC}(\mathbf{\Omega}, \mathbf{v}, t)  \right]}_{\hat{U}(\mathbf{\Omega}, \mathbf{v})}|{\bm \phi}\rangle \ .\label{eq:ansatz}
\end{eqnarray}
Here, $\hat{\mathcal{T}}$ denotes the time ordering operator, $T$ is the total pulse duration, and $|{\bm \phi}\rangle=\ket{\phi_1}\cdots\ket{\phi_N}$ denotes the initial state of the $N$ qubits. We will either start with all qubits in the $\ket{0}$ state or apply Pauli-$X$ gates to some of the qubits. The amplitudes and frequencies $\mathbf{\Omega}, \mathbf{v}$ are the parameters of the ansatz and will be optimized for a given objective function. In the following, we will refer to the unitary gate $\hat{U}(\mathbf{\Omega}, \mathbf{v})$ as the QOC ansatz.

\begin{figure}
    \centering
    \includegraphics[width=0.9\linewidth]{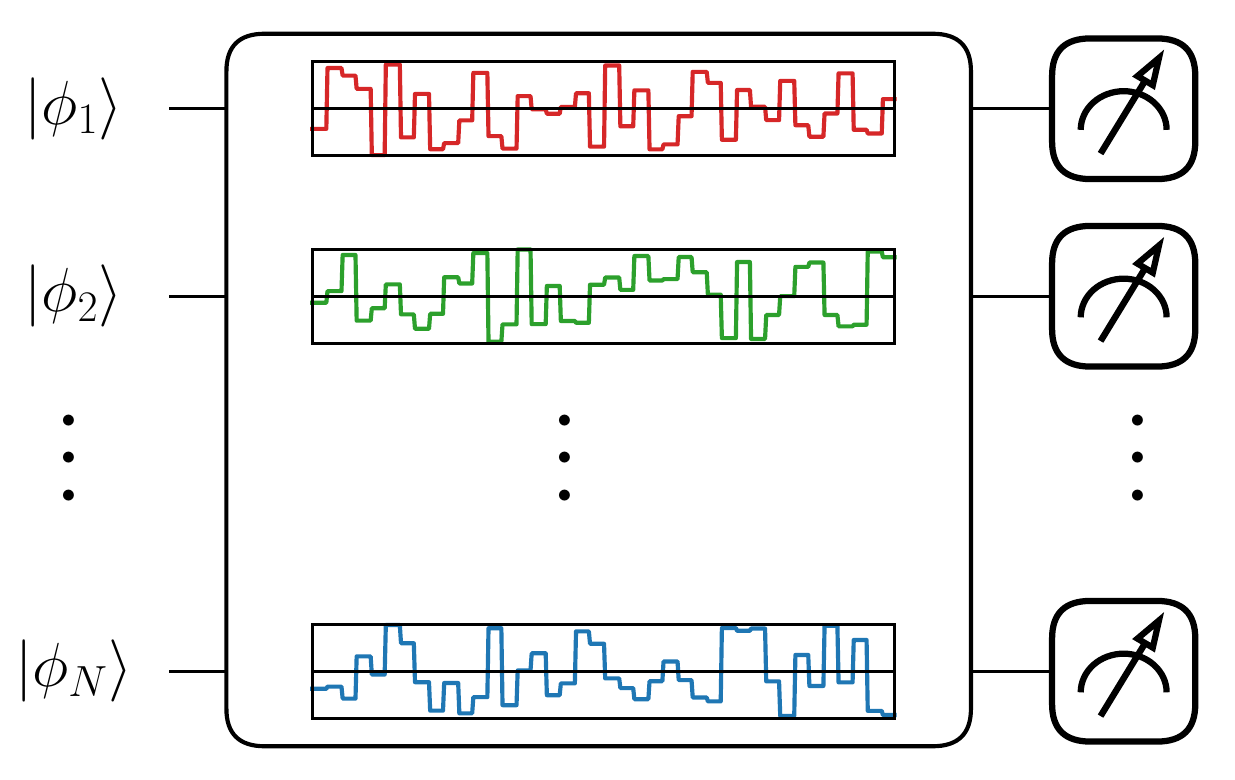}
    \caption{\it Schematic representation of the QOC ansatz. Each colored distribution represents the time-dependent pulse sequence applied to the corresponding qubit.}
    \label{fig:qoc}
\end{figure}

Fig.~\ref{fig:qoc} shows a schematic illustration of the QOC ansatz, where each colored pulse sequence is applied to a different qubit with initial state $\ket{\phi_i}$. Each panel shows the amplitude $\Omega_i(t)$ as a function of time for a given qubit.

\section{Ground state preparation with QOC}\label{sec:state-prep-qoc}

In this section, we describe the ground state preparation of the Schwinger model using the QOC ansatz described above. Using classical simulations, we quantify the potential speed-up using pulse-level optimization and the leakage to higher states, and we explore the dependence on device parameters and noise.

\subsection{Optimization algorithm}

Before discussing the pulse-based variational state preparation, we briefly describe the gate-based approach. VQAs involve a unitary operator that consists of a sequence of parameterized gates, which we denote by $\hat U(\Theta)$. Here, $\Theta$ denotes the trainable parameters of the ansatz. To prepare the ground state of a given Hamiltonian $\hat H$, these parameters are optimized with respect to the Hamiltonian's expectation value
\begin{eqnarray}
    \min_\theta \left\langle 0^{\otimes N} \big|\hat U^\dagger(\Theta) \hat{H} \hat U(\Theta)\big| 0^{\otimes N}\right\rangle \ .\label{eq:vqa}
\end{eqnarray}
Here, we start with all qubits in the $\ket{0}$ state. Once optimized, the Hamiltonian is approximately diagonalized by $\hat U(\Theta)$, yielding a ground state estimation $\hat U(\Theta)| 0^{\otimes N}\rangle$.

For the state preparation using QOC, the unitary operator $\hat U(\Theta)$ in Eq.~\eqref{eq:vqa} is replaced by the pulse-level ansatz $\hat{U}(\mathbf{\Omega}, \mathbf{v})$ given in  Eq.~\eqref{eq:ansatz}. The pulse amplitudes $\mathbf{\Omega}$ and frequencies $\mathbf{v}$ are optimized to approximate the ground state
\begin{eqnarray}\label{eq:minpulse}
    \min_{\mathbf{\Omega}, \mathbf{v}} \langle \Psi(\mathbf{\Omega}, \mathbf{v})| \hat{H} |\Psi(\mathbf{\Omega}, \mathbf{v})\rangle \ .
\end{eqnarray}
We apply $n$ pulses to each qubit $i$ with amplitude $\Omega_{i}\in\mathbf{\Omega}$ and phase $v_i\in \mathbf{v}$. We parametrize the amplitudes $\Omega_i(t)$ for each qubit as a piecewise constant function with $n$ segments 
\begin{eqnarray}
    \Omega_i(t) = \begin{cases}
        \Omega_{i,1}\ , & {\rm for\ } t\leq t_1,\\
        \Omega_{i,2}\ , & {\rm for\ } t_1 < t \leq t_2,\\
         & \vdots \\
        \Omega_{i,n}\ , & {\rm for\ } t_{n-1}< t \leq t_n
    \end{cases} \ .
    \label{eq:pwc}
\end{eqnarray}
Here, $t_n=T$ is the predetermined total pulse duration. Note that $\Delta t = t_n - t_{n-1} = T/n$ is highly dependent on the pulse resolution of the quantum device. We also considered variable pulse durations but observed undesirable outcomes, likely due to overparameterization. Instead, we obtain the minimum pulse duration using a search algorithm, which iteratively increases the pulse duration until the optimal duration is achieved. The pulse amplitudes are constrained to $\Omega_{i} \in 2\pi \times [-20,20]$~MHz, and each qubit's drive frequency is allowed to deviate from the transition frequency between the $\ket{0}$ and $\ket{1}$ state, $\omega_i$, by a maximum of $2\pi \times 1$~GHz, see for instance Ref.~\cite{Asthana:2022pgu}. We define this difference as $\Delta \nu_i=\omega_i-v_i$. The values of $\omega_i$ and $\delta_i$ used for our simulations are provided in Table~\ref{tab:omega-delta}. Additionally, the inter-qubit coupling values for the nearest-neighbor (all-to-all) architecture are provided in the top (bottom) panel of Table~\ref{tab:coup-transmon}.\footnote{The values in Tables~\ref{tab:omega-delta} and~\ref{tab:coup-transmon} are taken from Ref.~\cite{Meitei:2021aa} and referenced in the \texttt{ctrl-VQE} package, which can be found at \href{https://github.com/oimeitei/ctrlq/tree/7df76db0b3677447a2027c0a3a1176beb7267c45}{this GitHub page}, with commit ID \texttt{7df76db}.} We explore both topologies below.

\begin{table}[t]
    \centering
    \setlength\tabcolsep{8pt}
    \begin{tabular}{|c||c|c|c|c|}
        \hline
        $i$ & 1 & 2 & 3 & 4 \\\hline\hline
        $\omega_i/2\pi$ & 4.808 & 4.833 & 4.940 & 4.796\\
        $\delta_i/2\pi$ & 0.310 & 0.292 & 0.330 & 0.262 \\\hline
    \end{tabular}
    \caption{\it Parameters of the four transmon qubits used for our simulations given in angular frequencies, with units in GHz.}
    \label{tab:omega-delta}
\end{table}

\begin{table}
    \centering
    \setlength\tabcolsep{8pt}
    \begin{tabular}{|c|c|c|}
    \multicolumn{3}{c}{nearest-neighbor}\\
    \hline
         $1\leftrightarrow 2$ & $2\leftrightarrow 3$ & $3\leftrightarrow 4$\\\hline\hline
         18.3 & 21.3 & 19.3 \\\hline
    \end{tabular}
    \\\vspace{.25cm}
    \begin{tabular}{|c|c|c|c|c|c|}
    \multicolumn{6}{c}{all-to-all}\\
    \hline
         $1\leftrightarrow 2$ & $1\leftrightarrow 3$ & $1\leftrightarrow 4$ & $2\leftrightarrow3$  & $2\leftrightarrow 4$ & $3\leftrightarrow 4$ \\\hline\hline
         18.3 & 21.3 & 19.3 & 20.3 & 18.3 & 21.3 \\\hline
    \end{tabular}
    \caption{\it Inter-qubit coupling strengths $g_{ij}/2\pi$ for the nearest-neighbor (top panel) and the all-to-all coupling architecture (lower panel) in MHz.}
    \label{tab:coup-transmon}
\end{table}

\begin{figure*}
    \centering
    \includegraphics[width=0.85\textwidth]{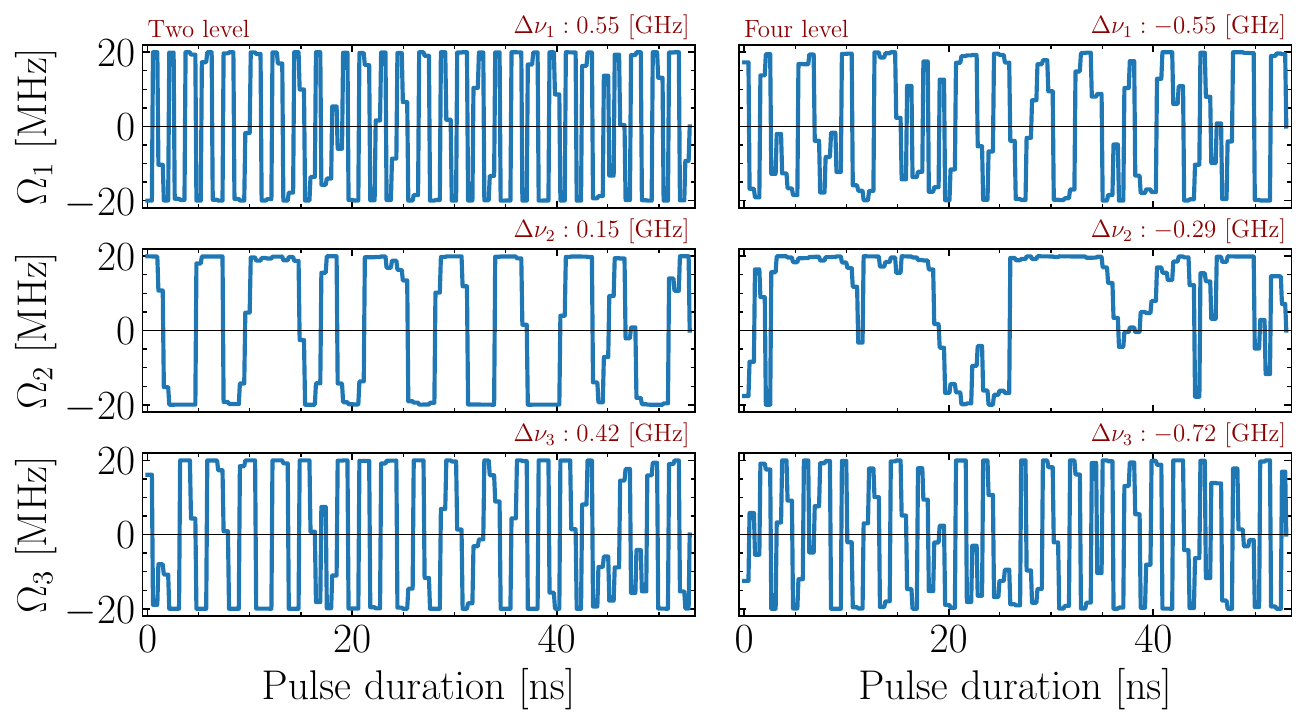}
    \caption{\it The left panels display the optimal pulse sequence obtained for a three-qubit Schwinger model where the transmon qubit is approximated as a two-level system. Each panel, from top to bottom, shows the pulse amplitude as a function of the pulse duration for each qubit. Additionally, changes in frequency are indicated at the top of each panel. The right panels show the same, but for the case where the transmon qubits are approximated as a four-level system. The results were obtained by initializing the circuit in the $|010\rangle$ state, and for both approaches, we found the MET to be 53 ns.~\label{fig:n3-pulse-nearestneig}}
\end{figure*}

We carry out the classical simulations using the \textsc{PennyLane} package (version 0.35.1)~\cite{bergholm2020pennylane} and \textsc{Jax} (version 0.4.23)~\cite{jax2018github}, with custom extensions for higher-dimensional states during the evolution of Eq.~\eqref{eq:ansatz}. For each qubit, we use 100 pulse segments (unless stated otherwise), resulting in $n+1$ trainable parameters per qubit, which includes the time-independent phase. Both pulse amplitudes and frequencies were optimized within the aforementioned limits\footnote{We used square pulses as shown in Fig.~\ref{fig:qoc}. We also tested flat-top Gaussian pulses, but their amplitude gradients were on the order of $\mathcal{O}(10^{-7})$, hindering an effective optimization.}. The gradient-based optimization was performed using \textsc{SciPy} (version 1.10.0)~\cite{2020SciPy-NMeth} with the \texttt{L-BFGS-B} algorithm. While we limit ourselves in this work to simulations where the gradient is obtained through differentiable programming. We note that the gradients of pulse programs can be obtained using stochastic parameter shift rules and analytic methods~\cite{Banchi_2021, Leng:2022ypv, Kottmann:2023xqg}.

A one-to-one comparison between gate-based and QOC-based ansätze is challenging due to the multitude of gate-based constructions available. To provide two opposite ends of the spectrum, we will compare QOC to both the Hamiltonian-based ansatz~\cite{Gard_2020} and the strongly entangling layer~\cite{Schuld:2018ahn}. The Hamiltonian-based ansatz involves exponentiating the target Hamiltonian, \( e^{-i\theta \hat{H}} \) by  performing a Trotter decomposition~\cite{BOGHOSIAN199830, doi:10.1126/science.273.5278.1073}. We can write the Hamiltonian as $\hat H=\sum_j \hat H_j$, where each term is given by a coefficient multiplied by a Pauli string $\{\mathds{1}, X, Y, Z\}^{\otimes {\rm dim}[\hat{H}]}$. A first-order Trotter decomposition can be written as
\begin{equation}\label{eq:trotter}
    e^{-i\theta\hat H} \approx \prod_j e^{-i\theta \hat{H}_j} \,,
\end{equation}
where the error depends on the non-commuting terms of the Hamiltonian and the value of $\theta$. Each factor of the trotterized formula can be mapped exactly to elementary quantum gates~\cite{doi:10.1126/science.273.5278.1073}. For the Hamiltonian-based ansatz, different trainable parameters are used in Eq.~\eqref{eq:trotter}, and multiple layers can be combined to increase the expressibility of the ansatz. The second ansatz we consider as a reference is built from strongly entangling layers. Here, each qubit is initiated with three consecutive Pauli rotation gates, \( R_Z(\theta_1) R_Y(\theta_2) R_Z(\theta_3) \). This is followed by CNOT gates that connect every neighboring qubit.

We note that throughout this paper, we compare the QOC approach exclusively to two specific gate-based ansätze. Depending on the complexity of the model, gate-based ansätze may require multiple layers to achieve the desired accuracy. Given the wide variety of available circuit designs, we restrict our comparison to a single layer of these chosen ansätze. This approach minimizes bias arising from the selection of circuit design and the challenges associated with the optimization procedure.

Furthermore, we would like to stress that the QOC algorithm inherently allows for significantly more trainable parameters compared to gate-based approaches, owing to its scalability. However, the large number of variational parameters can pose significant challenges for optimization algorithms, potentially leading to an increased classical computational overhead due to the additional number of steps required to minimize the loss function. We leave the exploration of this aspect for future work.

\subsection{Numerical results}

To assess the feasibility of using the variational QOC ansatz for the state preparation of lattice field theories, we start by considering the Schwinger model with a topological term described in section~\ref{sec:schwinger}. Due to the complexity inherent in certain Hamiltonians, deep variational circuits may be required to find the ground state accurately. By enabling lower-level control of the quantum device, QOC allows for more efficient ansätze, resulting in shorter pulse durations and reduced coherence-time requirements. For the Schwinger model, part of the difficulty arises due to the long-range interaction in Eq.~\eqref{eq:HZZ} that necessitates SWAP gates if the device's connectivity is limited. The ground state of the Schwinger model can be particularly challenging as it is given by a superposition of alternating fermion spin configurations.  

\begin{figure*}[t]
    \subfigure[Two-level system]{\includegraphics[width=0.42\linewidth]{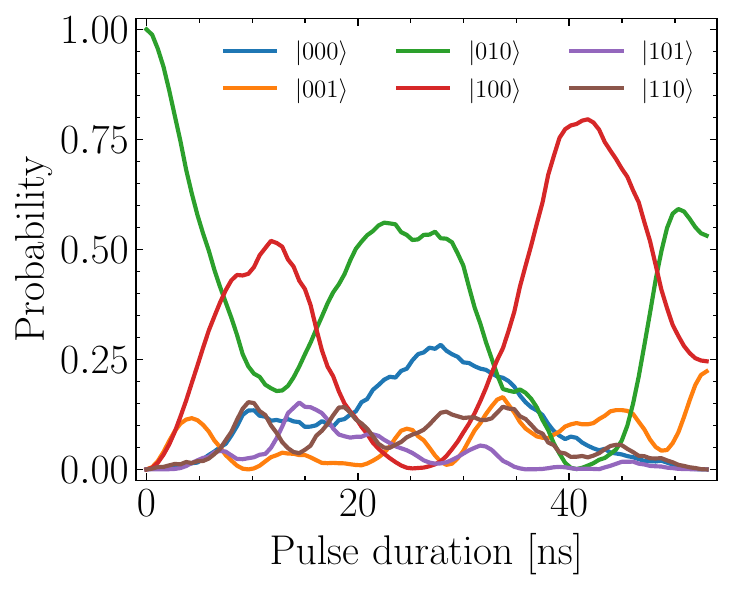}}\quad\quad\quad\quad
    \subfigure[Four-level system]{\includegraphics[width=0.42\linewidth]{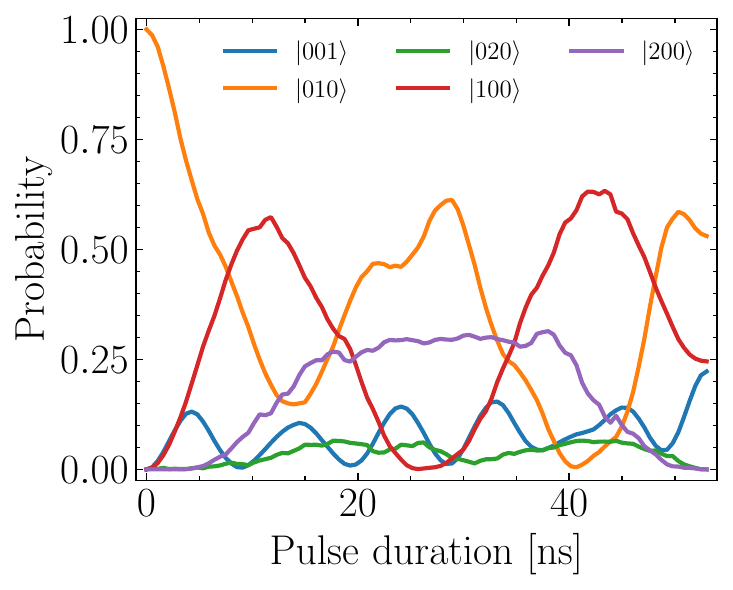}}
    \caption{\it The left panel shows the probability of each state with at least 10\% maximum probability for the optimal pulse sequence shown in Fig.~\ref{fig:n3-pulse-nearestneig}, left panels. The right panel shows the same but for the four-level approximation of the transmon qubits, corresponding to the right panels of Fig.~\ref{fig:n3-pulse-nearestneig}.}
    \label{fig:n3_prob}
\end{figure*}

As a first step, we variationally prepare the ground state of the three-site or three-qubit Schwinger model using the QOC ansatz described above. We choose exemplary values for the fermion mass $m=0.5$, the lattice spacing $a=0.1$, the topological angle $\theta=0.5$, and the electric charge $e=0.2$. The three-qubit Schwinger model serves as an ideal initial testbed as it still captures essential features of the Schwinger model. In particular, its ground state is a superposition of multiple states. Using exact diagonalization, we find that the ground state is given by a superposition of three states
\begin{eqnarray}
    |\Psi_{\rm 3-site}\rangle = 0.223 |001\rangle + 0.531 |010\rangle + 0.246 |100\rangle .\ \ \label{eq:schwinger-3-site-gs}
\end{eqnarray}
Using the QOC specifications detailed in section~\ref{sec:state-prep-qoc}, we optimize the pulses applied to the three qubits using the nearest-neighbor qubit architecture. Throughout this work, we will refer to the minimum pulse duration or MET as the minimum time at which the solution can be reached with the desired accuracy using the search algorithm described above. For the three-site case, we obtained a MET of \(53.0 \pm 0.5\) ns, where the error indicates the resolution of the search algorithm. Fig.~\ref{fig:n3-pulse-nearestneig} shows the pulse amplitudes $\Omega_i(t)$ and frequency deviations $\Delta \nu_i$ for each qubit. The left panels show the result when the infinite-dimensional Hilbert space of the coupled anharmonic oscillators in Eqs.~\eqref{eq:drive} and~\eqref{eq:control} are truncated to two levels, i.e. only the states $\ket{0}$ and $\ket{1}$ are included. Instead, the right panels show the four-level truncation results where the states $\ket{2}$ and $\ket{3}$ are also included, which we describe in more detail below. The panels show the pulse sequence applied to each of the three qubits. The pulse shapes predominantly exhibit a so-called bang-bang form where the constrained control parameters saturate their bounds. This is consistent with the findings in optimal control theory, indicating that the variationally obtained pulses closely approximate the optimal solution. We refer the reader to Refs.~\cite{Moore_Tibbetts_2012, Yang_2017,Lin_2019, Poggi:2020ibw, Brady_2021, PhysRevApplied.19.064071, Asthana:2022pgu} for more details. For both the two- and four-level systems, we achieve a ground state energy that differs from the result using exact diagonalization \(\Delta E = \lVert E_{\rm truth} - E_{\rm reco} \rVert\) by less than \(10^{-3}\). Notably, we were only able to achieve this short pulse duration when the qubits were initialized in the mass eigenstate of the Schwinger model, which is given by \(|010\rangle\). Applying a Pauli-$X$ gate to one of the qubits takes approximately an additional 71 ns for our setup. Therefore, the total time to prepare the ground state of the three-site Schwinger model is 124 ns.

In order to compare the achieved MET to gate-level approaches, we consider the two gate-level variational approaches described above. Constructing a single-layer Hamiltonian-based ansatz for the three-site Schwinger model requires $34$ gates, with a circuit depth of $24$, including $10$ CNOT gates. Depending on the quantum hardware, native two-qubit gates include, for example, CNOT, CZ, or Echo Cross Resonance (ECR) gates. The device parameters in tables~\ref{tab:omega-delta} and~\ref{tab:coup-transmon} are representative values for IBM's Falcon processors taken from Ref.~\cite{Meitei:2021aa} where the \texttt{ctrl-VQE} approach was introduced. In this case, CNOTs are the native two-qubit gates. As a representative value for the CNOT gate time, we assume $400$~ns and $71$~ns for single-qubit gates. A back-of-the-envelope calculation yields an execution time of $7$~$\mu$s for a single Trotter step layer. This corresponds to a $40\times$ speed up when using the QOC ansatz. In contrast, a strongly entangling layer requires $12$ gates with a depth of $6$, including $3$ CNOT gates. Executing such a circuit requires $2$~$\mu$s, corresponding to a QOC speedup of $11\times$. These rough comparisons indicate significantly reduced coherence time requirements when using the QOC ansatz.

\begin{figure*}[t]
    \centering
    \subfigure[First half]{\includegraphics[width=0.9\textwidth]{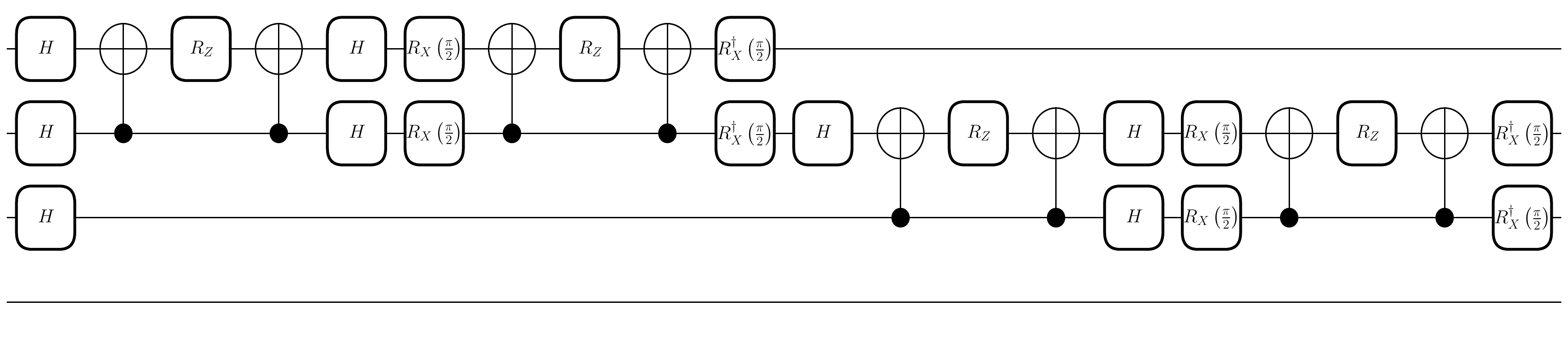}\label{fig:schwinger_trota}}
    \subfigure[Second half]{\includegraphics[width=0.9\textwidth]{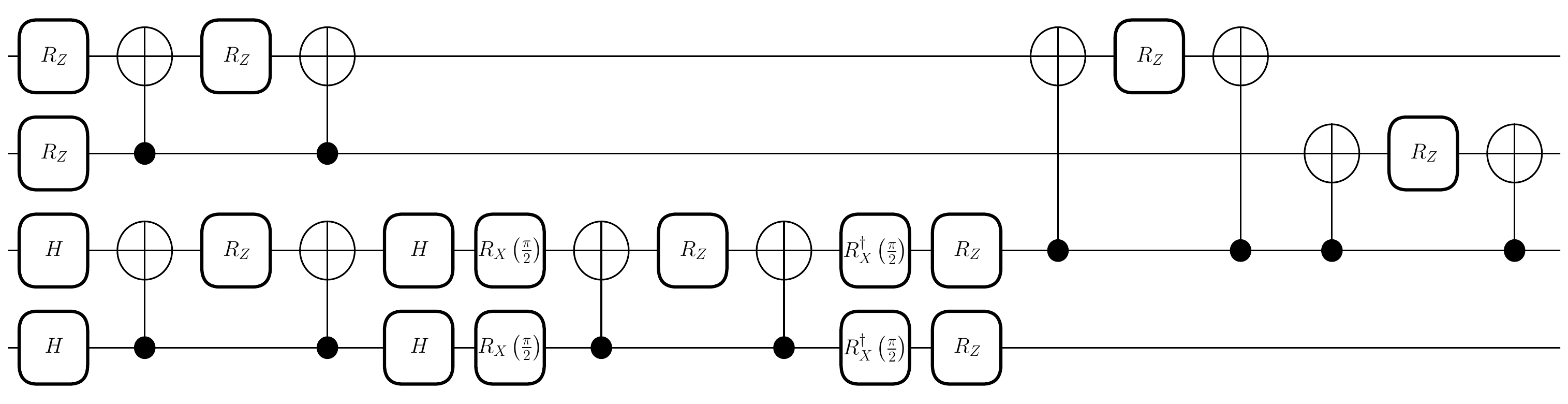}\label{fig:schwinger_trotb}}
    \caption{\it Gate-based representation of a single Trotter step of the four-site Schwinger model. The rotation of each $R_Z$ gate depends on the parameters in the Hamiltonian and Trotter step size.}
    \label{fig:schwinger_trot}
\end{figure*}

As mentioned above, the right panels of Fig.~\ref{fig:n3-pulse-nearestneig} show the result when we allow for leakage to higher-level states in the simulation. While leakage can occur during the evolution, the measurement is only sensitive to the lowest two states. We observed that no leakage occurs to the state $\ket{3}$, ensuring that no information is lost due to yet higher states that are not included in our simulation. We also note that the authors of Ref.~\cite{PhysRevApplied.19.064071} observed a shorter pulse duration when leakage to higher states is allowed at the end of the pulse sequence. If a higher-level state is obtained, it is discarded and compensated for by increasing the number of shots. We observed similar results, but the reconstruction accuracy degraded in our case. This is likely due to the fact that the ground state here is given by a superposition of three states with roughly equal weights, see Eq.~\eqref{eq:schwinger-3-site-gs}. Fig.~\ref{fig:n3_prob} shows the probability for the different stages during the evolution for the two and four-level truncation of the transmon Hamiltonian. While we observe a different path when leakage is included in our simulation, we are able to successfully reconstruct the ground state in Eq.~\eqref{eq:schwinger-3-site-gs} in both cases. Both simulations have the same total pulse duration, consistent with the findings in Ref.~\cite{PhysRevApplied.19.064071} when leakage only occurs during the evolution but not at the final time $T$. We observe a leakage of at most 30\% to the state \(|200\rangle\) between 5 ns to 40 ns (purple line in the right panel), and 5\% leakage to \(|020\rangle\) (green line in the right panel) for the same time interval. The leakage to both of these states subsides toward the end of the evolution, and we do not observe any leakage to the highest state $\ket{3}$.

Next, we turn to the simulation of the four-site Schwinger model. The asymmetric long-range interaction in Eq.~\eqref{eq:HZZ} only starts contributing to four sites, making the ground state preparation more challenging for both gate and pulse-level optimization. Using exact diagonalization, we find that the ground state is given by a superposition of six states
\begin{eqnarray}
     |\Psi_{\rm 4-site}\rangle &=& 0.055 |0011\rangle + 0.299 |0101\rangle + 0.202 |0110\rangle \nonumber\\
     &+& 0.202 |1001\rangle + 0.205 |1010\rangle + 0.038 |1100\rangle \ .\nonumber \\
\end{eqnarray}
To illustrate the additional challenge due to the long-range interaction, we note that implementing a single Trotter step for the four-site Schwinger model requires 55 different gates, and only a subset of these can be executed in parallel. Fig.~\ref{fig:schwinger_trot} shows the quantum circuit needed to perform a single Trotter step of the Schwinger model. Note that although Fig.~\ref{fig:schwinger_trota} includes only nearest-neighbor inter-qubit couplings, Fig.~\ref{fig:schwinger_trotb} includes two CNOT gates that connect the first and third qubit requiring two additional SWAP gates to rearrange the qubits. Assuming again 400 ns for the CNOT gate time and 71~ns for single-qubit gates, excluding the SWAP gates, the circuit shown in Fig.~\ref{fig:schwinger_trot} takes approximately $7.9~\mu$s to execute.

\begin{figure*}[t]
    \centering
    \subfigure[Nearest-neighbor coupled qubit architecture, MET 181 ns.]{\includegraphics[width=0.9\linewidth]{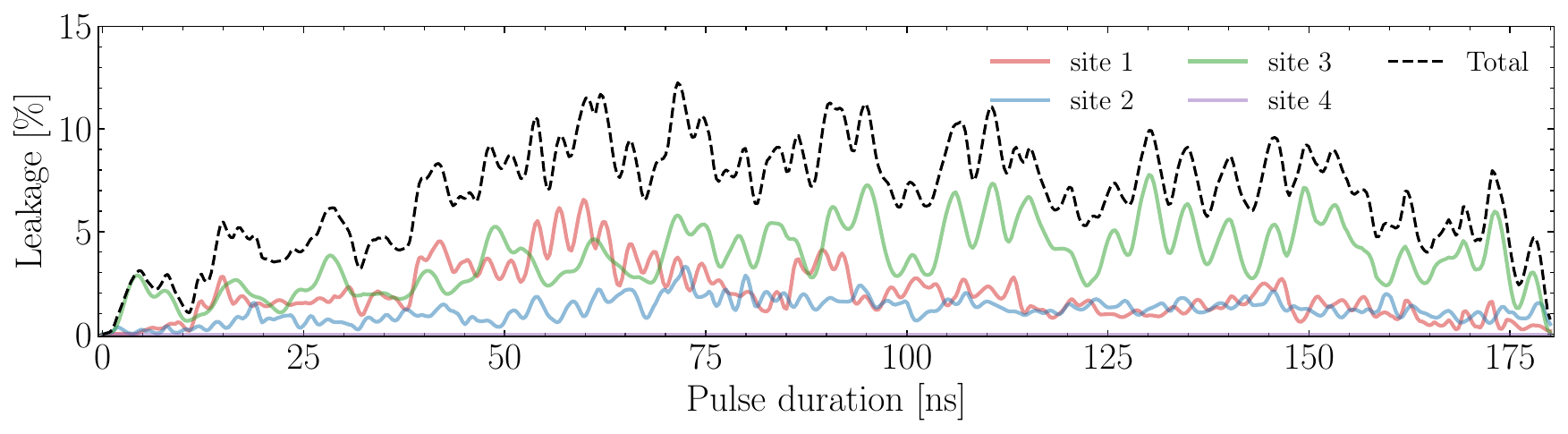}\label{fig:n4_nearest_neig}}
    \subfigure[All-to-all connected qubit architecture, MET 101 ns.]{\includegraphics[width=0.9\linewidth]{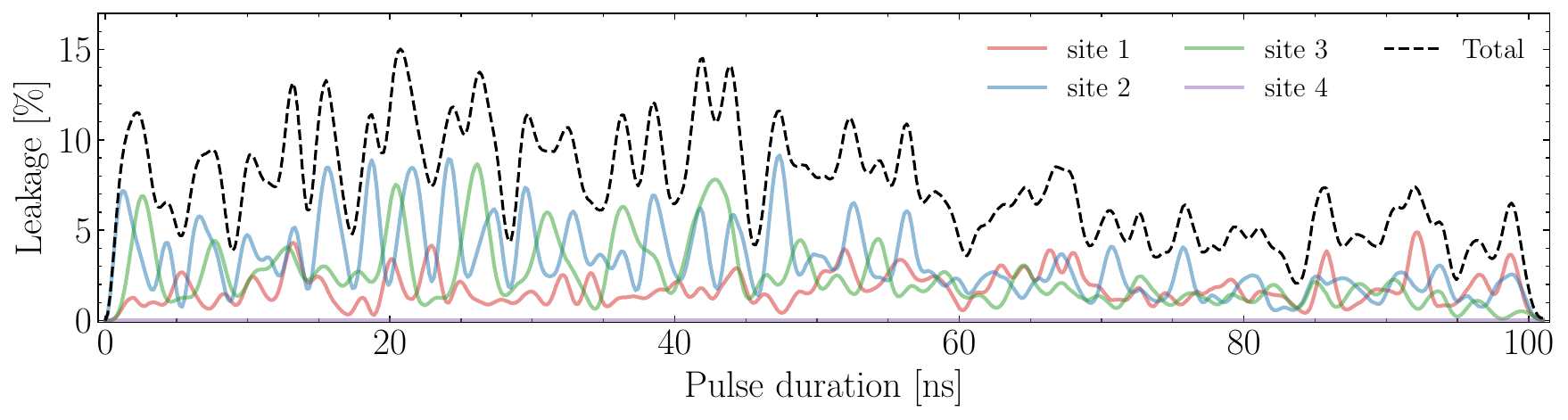}\label{fig:n4_all-to-all}}
    \caption{\it Leakage to higher-level states for different qubit architectures. The solid lines show the leakage for each qubit where red, blue, green, and purple correspond to the first, second, third, and fourth qubit. The dashed black line shows the total leakage.}
    \label{fig:n4_archi_prob}
\end{figure*}

We compare this result to the QOC-based ground state preparation using the nearest-neighbor and all-to-all qubit architectures. Given the long-range interaction term, we expect the difference between the two architectures to be particularly pronounced for the four-site Schwinger model. Fig.~\ref{fig:n4_archi_prob} shows the leakage to higher-level states as a percentage for each site or qubit (colored solid lines) as well as the total leakage (dashed black line) as a function of the pulse duration. The two panels show the result for the two-qubit architectures with nearest-neighbor (Fig.~\ref{fig:n4_nearest_neig}) and all-to-all (Fig.~\ref{fig:n4_all-to-all}) connectivity. In both cases, we achieved an inaccuracy of \(\Delta E \lesssim \mathcal{O}(10^{-3})\). The obtained MET values for the two architectures are given by 181 ns with only nearest-neighbor couplings and 101 ns for the all-to-all architecture. Using the QOC-ansatz, we find a speedup of $43\times$ compared to a single layer of a Hamiltonian-based ansatz or Trotter step for the nearest-neighbor architecture. The speedup is even more significant for the all-to-all architecture, where the obtained speedup increases by yet another factor of more than 2. This is consistent with the expectation that the long-range interaction allows for significant gains using QOC. We note that for the QOC-based ground state preparation, we initialized the qubits again in the mass eigenstate of the Schwinger model \(|0101\rangle\). Preparing this state requires two Pauli-$X$ gates, which are taken into account for the assessment of the achieved improvement. In both simulations, we observed less than 15\% leakage in any of the qubits. Although our leakage estimation includes both the states \(|2\rangle\) and \(|3\rangle\) for each qubit, we did not observe any leakage to the \(|3\rangle\) states. This limited leakage is due to the combination of the magnitude of anharmonicity in the transmon Hamiltonian (see Table~\ref{tab:coup-transmon}) and the constraints on the pulse amplitude, which does not provide enough energy for the state to leak into the state \(|3\rangle\).

\subsection{Algorithmic constraints}

\begin{figure}[t]
    \centering
    \includegraphics[width=0.9\linewidth]{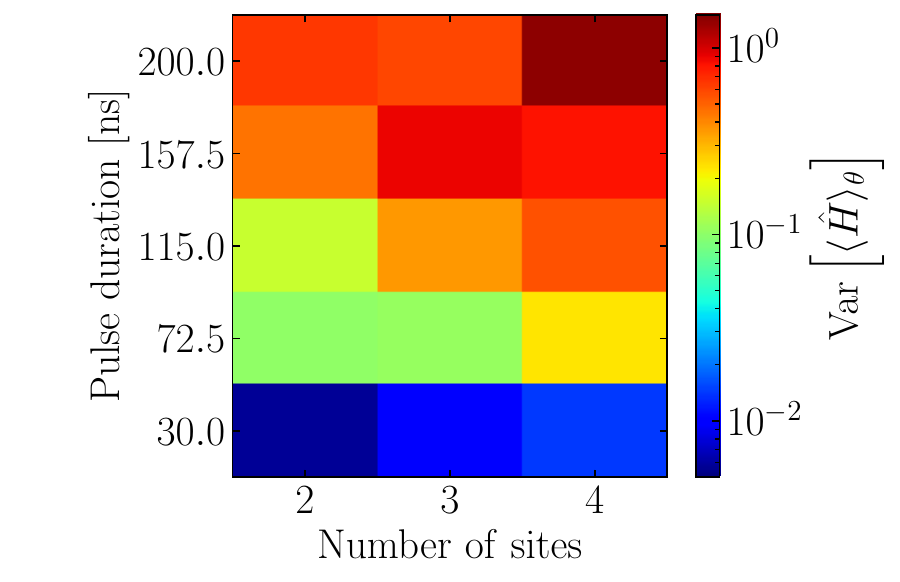}
    \caption{\it Change in the variance of expectation value with respect to pulse duration and number of sites. The variance has been estimated by initiating the circuit with randomly sampled 100 parameters $\mathbf{\Omega}, \mathbf{v} \in \theta$.}
    \label{fig:variance}
\end{figure}

Gate-based VQAs often suffer from vanishing gradients, known as barren plateaus~\cite{McClean:2018jps, Ragone:2023qbn} and local traps~\cite{PhysRevLett.127.120502}. This can limit the scalability of the variational algorithm in terms of both the number of qubits and the depth of the circuit. In analogy to the gate-based approach, we assess potential limitations of the QOC setup by considering the variance of the expectation value of the Schwinger model Hamiltonian for different pulse durations and numbers of sites. Fig.~\ref{fig:variance} shows the change in variance with respect to both of these parameters.

For this estimation, we fixed the number of pulse segments to 100 per qubit and sampled both the qubit amplitudes and phases from a uniform distribution within the constraints stated in section~\ref{sec:state-prep-qoc}. Each grid point shows the variance for 100 independent samples, and the number of parameters scales as \(N \times 101\), where \(N\) is the number of qubits. We observe that larger pulse resolutions lead to larger values of the variance, indicating that we do not obtain a flat loss function for the system sizes that we were able to explore. Interestingly, we also observe that the variance increases as the number of qubits is increased. This suggests a favorable landscape for the optimization even for a large number of qubits. We conclude that the QOC ansatz allows for a relatively large number of parameters, which increases the expressibility of the ansatz. However, it is essential to note that this may also lead to overparameterization, which needs to be explored in more detail in future work. Currently, extending our studies to a larger number of qubits is limited by our computing resources.

\subsection{Hardware constraints}

\begin{figure}
    \centering
    \includegraphics[width=0.9\linewidth]{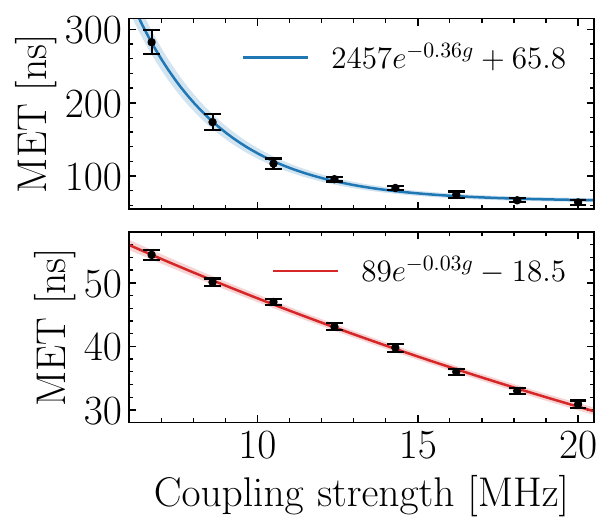}
    \caption{\it Scan of the achieved MET as a function of the inter-qubit coupling strength \(g_{ij}\) of the transmon Hamiltonian, see Eq.~\eqref{eq:drive}, for the nearest-neighbor architecture. The upper panel shows the obtained MET values for the three-site Schwinger model, while the lower panel shows the MET values for the two-site case. The black markers show the results obtained from QOC, along with the optimization error. The solid lines represent the best fit to the mean QOC results, and the shaded area represents the error of the fit. The circuits in the upper (lower) panel are initialized as \(|010\rangle\) (\(|10\rangle\)).}
    \label{fig:coup-met}
\end{figure}

Beyond the architecture and ability to control the transmon Hamiltonian, the strength of the inter-qubit interaction significantly impacts the efficiency of the variational algorithms. To estimate the effect of the inter-qubit coupling strength in the transmon Hamiltonian, \( g_{ij} \) in Eq.~\eqref{eq:drive}, we fix \( \omega_i \) and \( \delta_i \) to the values presented in Table~\ref{tab:omega-delta}. We then determine the MET for different values of \( g_{ij} \) for the two-site and three-site Schwinger models, as depicted in Fig.~\ref{fig:coup-met}. 

The lower panel of Fig.~\ref{fig:coup-met} shows the results for the two-qubit QOC ansatz. Here, the black markers represent the mean of ten different results, and the error bars indicate the square root of the optimization variance, i.e., the standard deviation of ten successive results and scan resolution to find the MET. The solid red curve and the shaded area around it represent the best fit and its error. Similarly, the upper panel displays the same for the three-site Schwinger model. For the estimation of the MET, we required each successful point to achieve \( \Delta E \lesssim \mathcal{O}(10^{-3}) \). As before, each point was executed for an initial state prepared in the mass eigenstate of the Schwinger model. This additional time of 71 ns is not included in the figure as it is only a constant offset.

Our results indicate that an exponential improvement of the MET can be achieved over a certain range of the inter-qubit coupling strength. This improvement becomes more pronounced with larger system sizes. We observe much sharper gradient values for our best-fit curve for the three-qubit compared to the two-qubit results. Additionally, the optimization error steadily increases with the decrease in coupling strength. However, eventually, the improvements reach a plateau, as can be seen in the upper panel of Fig.~\ref{fig:coup-met}. 

Next, we will investigate the MET values that may be achieved with the currently publicly available devices from IBM. Unlike the Falcon devices above, the native two-qubit operations on these Eagle processors are ECR gates. For the time estimates presented in this section, we account for the conversion of ECR to CNOT gates. We use the parameter values of the transmon Hamiltonian that are provided for \texttt{ibm\_osaka}, which are listed in Table~\ref{tab:ibmq-transmon}. Although the \(\omega_i\) and \(\delta_i\) values are similar to our previous simulations, the inter-qubit coupling strength is roughly an order of magnitude smaller compared to the values listed in Table~\ref{tab:coup-transmon}. We initiated the MET search scan for the two-qubit Schwinger model for ten different instances and only accepted \(\Delta E \lesssim \mathcal{O}(10^{-3})\). Our simulation converged to:
\[
    {\rm MET}_{\texttt{ibm\_osaka}} : 882.7 \pm 5.5_\text{resolution} \pm 8.7_\text{SEM}\ {\rm ns}\, .
\]
The uncertainty indicates the optimization error and the scan resolution, which is reported along with the standard error of the mean (SEM). Note that we assumed a pulse resolution of \(2.21 \pm 0.07\) ns for this simulation, and device noise has not been considered. 

At the gate level, a single layer of a Hamiltonian-based ansatz for the two-site Schwinger model includes 12 single-qubit gates, of which only seven cannot be executed in parallel, and 4 CNOT gates. The time to execute a two-qubit gate takes 660 ns, whereas a single-qubit gate takes 71 ns to execute on \texttt{ibm\_osaka} according to the IBM Quantum System information webpage. Therefore, a single Trotter step would take around \(3.14 \ \mu s\). This indicates a $4.2\pm 0.1\times$ improvement using QOC techniques, including the initial state preparation. For such a small system, it is also possible to use a single strongly entangling layer~\cite{Schuld:2018ahn} to find the ground state. This includes six single-qubit gates, with three that cannot be executed in parallel, and 2 CNOT gates, bringing the total duration of the layer to \(1.53 \ \mu s\). This indicates a $1.61 \pm 0.05 \times$ improvement, including the initial state preparation for the QOC ansatz. We note that this result also reveals the naïvetè of our best fit shown in Fig.~\ref{fig:coup-met}, where we estimate the MET for \texttt{ibm\_osaka} to be around 65 ns. Even though the values for \(\omega_i\) and \(\delta_i\) are comparable to the actual device in Table~\ref{tab:ibmq-transmon}, this result shows that the growth in MET with respect to the reduction in coupling strength is higher than estimated in Fig.~\ref{fig:coup-met}.

\begin{table*}[t]
    \centering
    \setlength\tabcolsep{9pt}
    \begin{tabular}{|c||c|c|c|c|c|c|c|c|}
    \hline
    & \multicolumn{2}{c|}{\texttt{ibm\_osaka}} & \multicolumn{2}{c|}{\texttt{ibm\_brisbane}} & \multicolumn{2}{c|}{\texttt{ibm\_sherbrooke}} & \multicolumn{2}{c|}{\texttt{ibm\_kyoto}} \\\hline
         $i$ & $1$ & $2$ & $1$ & $2$  & $1$ & $2$ & $1$ & $2$ \\\hline\hline
         $\omega_i/2\pi$ [GHz] & $4.977$ & $4.928$ & $4.878$ & $4.970$ & $4.792$ & $4.893$ & $5.063$ & $4.856$\\\hline
         $\delta_i/2\pi$ [GHz] & $0.309$ & $0.310$ & $0.312$ & $0.310$ & $0.313$ & $0.313$ & $0.308$ & $0$\\\hline
         $g_{1,2}/2\pi$ [MHz] & \multicolumn{2}{c|}{$2.03$} &  \multicolumn{2}{c|}{$2.03$}  &  \multicolumn{2}{c|}{$2.00$} &  \multicolumn{2}{c|}{$2.31$} \\\hline
    \end{tabular}
    \caption{\it Transmon Hamiltonian parameter values of the publicly available IBM quantum devices.}
    \label{tab:ibmq-transmon}
\end{table*}

We also tested the QOC algorithm using the parameters provided for \texttt{ibm\_brisbane} and \texttt{ibm\_sherbrooke}. Due to the significantly longer simulation time, we present our results for nine and five points, respectively. For these devices, we observed a substantially longer pulse duration to achieve the same order of accuracy for the ground state energy:
\[
    {\rm MET}_{\texttt{ibm\_brisbane}} : 1.88 \pm 0.01_{\rm  resolution}\ \pm 0.19_{\rm SEM}\ \mu{\rm s}
\]
with a pulse resolution of \(4.69 \pm 1.48\) ns and
\[
    {\rm MET}_{\texttt{ibm\_sherbrooke}} : 1.73 \pm 0.01_{\rm resolution}\pm 0.16_{\rm SEM}\ \mu{\rm s}
\]
with a pulse resolution of \(4.33 \pm 0.89\) ns. We note that due to limitations in computational time, the statistical power for determining the uncertainties for \texttt{ibm\_sherbrooke} is less than the other estimations. We estimate the improvement for \texttt{ibm\_brisbane} (\texttt{ibm\_sherbrooke}) compared to a single Trotter step as \(2.0 \pm 0.5\)$\times$ (\(2.1 \pm 0.3\)$\times$) and \(1.0 \pm 0.2\)$\times$ (\(1.0 \pm 0.27\)$\times$) for a single strongly entangling layer. For \texttt{ibm\_sherbrooke}, the median ECR gate execution time is 533.33 ns, which is reflected in our estimations. We note that a quantitative comparison between pulse and gate-level optimization is challenging since various aspects of the quantum hardware platforms can play an important role that may not be fully accounted for in our studies. That being said, we observe that the inter-qubit coupling strength, the complexity of the ground state, and the number of qubits play an important role.
Moreover, further improvements may be obtained using an additional two-qubit drive Hamiltonian as proposed in Ref.~\cite{Entin:2024jmy}. We leave more detailed studies, as well as quantum hardware simulations, for future work. As a first step toward simulations using quantum hardware, we will explore the optimization in the presence of noise in the following subsection. 

\subsection{Optimization in the presence of noise}

Beyond hardware constraints, such as the inter-qubit coupling strength, examining the effect of hardware noise on the optimization outcome is essential. Variational algorithms are usually resilient to noise as they can learn the structure of the noise during training~\cite{Dalton:2024aa, PhysRevA.109.042413, PhysRevA.109.032420}.

To assess the resilience of the QOC ansatz to noise, we employ \textsc{Qiskit} (version 1.0.2)~\cite{qiskit2024} and \textsc{Qiskit Dynamics} (version 0.5.1)~\cite{qiskit_dynamics_2023}. Although we use the same Hamiltonian in Eqs.~\eqref{eq:drive} and~\eqref{eq:control} to describe the coupled transmon qubits, we account for decoherence effects due to hardware noise by using a Lindblad or GKSL equation~\cite{Lindblad:1975ef,10.1093/acprof:oso/9780199213900.001.0001, Houdayer_2024, Brasil_2013, Andersson_2007}, which is given by
\begin{eqnarray}
    \frac{{\rm d} \rho}{{\rm d} t}  &=& -i [\hat{H}_{\rm OC}(\mathbf{\Omega}, \mathbf{v}, t),\ \rho] \label{eq:schrodinger-density}\\
    &+&\sum_i \left( 2\hat{L}_i \rho\hat{L}_i^\dagger - \left\{\hat{L}_i^\dagger\hat{L}_i,\ \rho \right\} \right)\ . \label{eq:lindblad}
\end{eqnarray}
The Lindblad equation describes the Markovian time evolution of the transmon qubits, an open quantum system in the limit of a sufficiently weak coupling to the environment. The first part of Eq.~\eqref{eq:schrodinger-density} is the time-dependent Liouville--von Neumann equation. It is analogous to the Schrödinger equation in Eq.~\eqref{eq:schrodinger}, but the time evolution is written in terms of the density matrix $\rho = |\Psi(t)\rangle \langle \Psi(t) |$. The second part, Eq.~\eqref{eq:lindblad}, extends the time-dependent Schrödinger equation to include the Lindblad operators $\hat{L}_i$. We account for amplitude damping, which is described by $\hat{L}_1 \equiv \sqrt{\Gamma_1}\ \hat{a}$ and dephasing which corresponds to the operator $\hat{L}_2 \equiv \sqrt{\Gamma_2}\ \hat{a}^\dagger\hat{a}$~\cite{Busel:2023gbg}. Here, $\Gamma_i$ is the rate of the state collapse due to damping or dephasing. For further details, we refer the reader to Ref.~\cite{Manzano_2020}\footnote{Similar tutorials can be found in the \href{https://qiskit-extensions.github.io/qiskit-dynamics/tutorials/Rabi_oscillations.html}{\textsc{Qiskit Dynamics} documentation v0.5.1}.}. Gate-to-pulse miscalibration has also been taken into account in our noise estimation\footnote{For details, we refer the reader to \href{https://learn.microsoft.com/en-us/azure/quantum/concepts-pauli-measurements}{the Microsoft Quantum article} on Pauli measurements and \href{https://qiskit-extensions.github.io/qiskit-experiments/tutorials/calibrations.html}{\textsc{Qiskit Dynamics} documentation for gate calibration.}}. 
Note that for simplicity, we avoid the use of additional optimizable two-qubit control terms in the transmon Hamiltonian. These enable the drive of a qubit at a rate that is proportional to the adjacent qubit frequency~\cite{Entin:2024jmy}. Such terms might reduce the sensitivity of the QOC to different initial states. We leave more detailed studies for future work.

For this analysis, we employ the parameterization provided for \texttt{ibm\_kyoto}, given in Table~\ref{tab:ibmq-transmon}. The corresponding state collapse rates can be found on the IBM Systems website. At the time of this analysis, $1/\Gamma_{1,2}$ for the first (second) qubit was given as $203.4\ \mu$s and $25.8\ \mu$s ($171.9\ \mu$s and $77.6\ \mu$s), respectively.

\begin{figure}[t]
    \centering
    \includegraphics[width=0.9\linewidth]{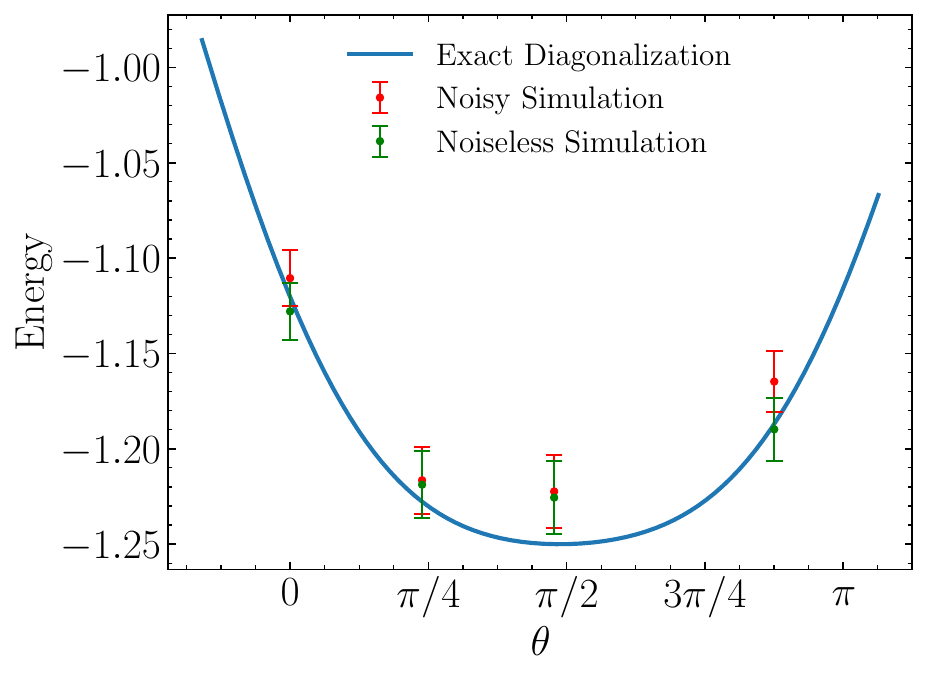}
    \caption{{\it The ground state energy as a function of the topological angle, plotted for exact diagonalization (solid blue line), noiseless simulation (green markers), and noisy simulation (red markers) where the error bars represent the standard deviation. The simulation was carried out for the specification of} \texttt{ibm\_kyoto}, see table~\ref{tab:ibmq-transmon}.}
    \label{fig:ibm_noise}
\end{figure}

\begin{figure*}
    \centering
    \includegraphics[width=0.75\linewidth]{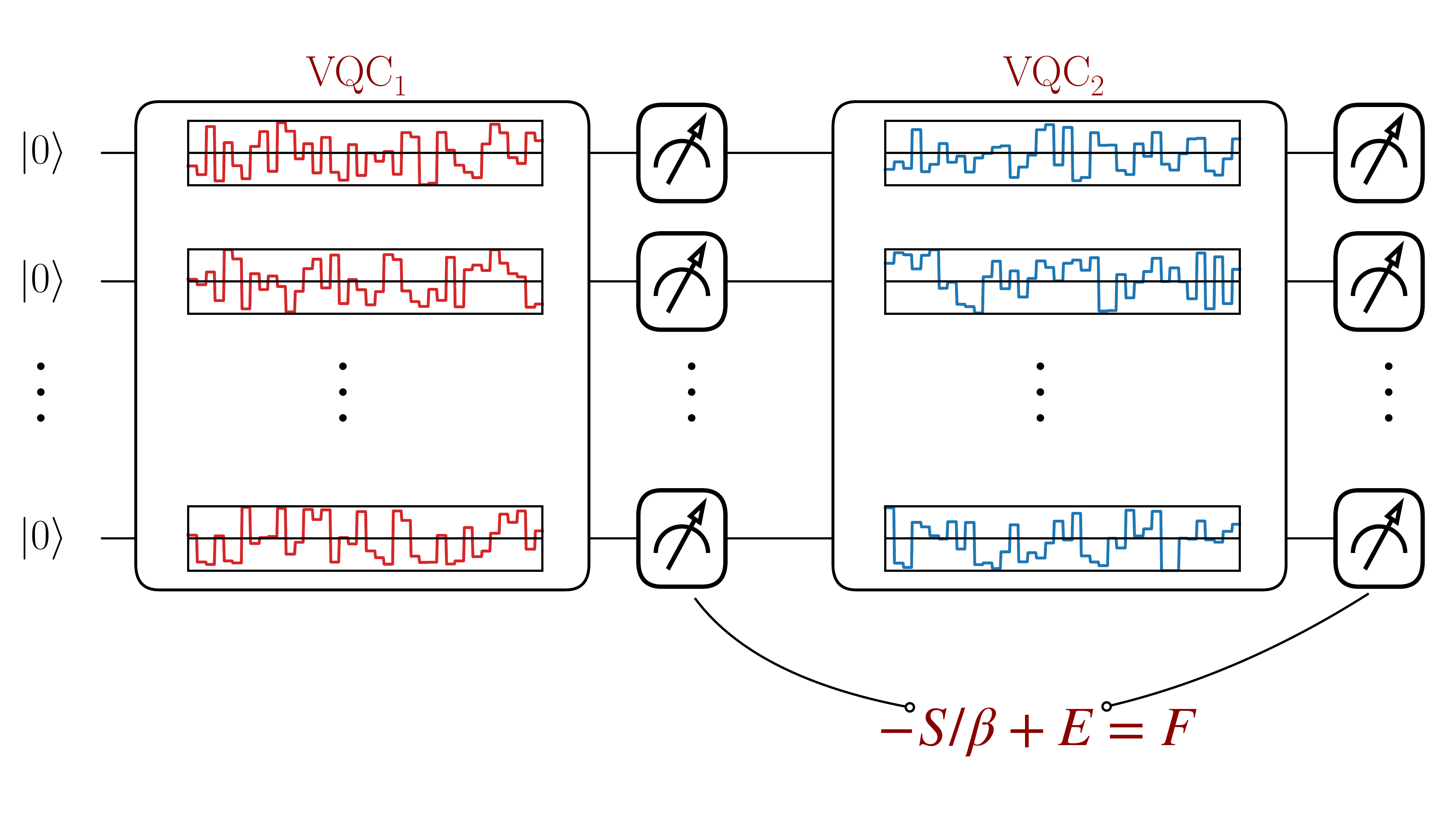}
    \caption{\it Schematic representation of the pulse-level VQT algorithm. The left block shows the first variational circuit (VQC$_1$), where the intermediate measurement in the computational basis is used to determine the entropy of the system. The right block takes the intermediate measurement as the initial state, and the output of the second variational circuit (VQC$_2$) is used to determine the energy of the ensemble.} 
    \label{fig:pulse_vqt_scheme}
\end{figure*}

To enhance the convergence rate of the optimization algorithm, we use individual phases for each pulse amplitude and limit the number of pulses to 70. This results in a total of 280 trainable parameters for a two-qubit system. Due to computational complexity, we limited the total pulse duration to 70 ns, and the initial state is set to $|00\rangle$. We ran our algorithm to find the ground state of the two-site Schwinger Hamiltonian with a lattice spacing of $a=0.5$ and four different values for the topological angle $\theta$. 

Fig.~\ref{fig:ibm_noise} shows our results for the ground state energy as a function of the $\theta$ angle. The blue curve shows the results using exact diagonalization, the green markers show the optimization results without noise (i.e., without Eq.~\eqref{eq:lindblad}), and the red markers show the results with noise. The error on each marker represents the standard deviation for 8192 shots. Our results show that the deviation due to noise is minimal. The overall deviation from the blue curve is due to the relation between the initial state and the fixed pulse duration. We did not attempt to find the optimal pulse duration to achieve a high reconstruction rate. Nonetheless, we achieved an inaccuracy of approximately $\Delta E \propto \mathcal{O}(10^{-2})$ with and without noise.

\section{Thermal state preparation with QOC}\label{sec:thermal-state}

In addition to preparing the vacuum ground state, we extend the QOC approach to finite temperatures and explore the preparation of thermal states of the Schwinger model. Different from the vacuum ground state, which involves a superposition of different pure states, the thermal state is described by an ensemble of states. It is described by the density matrix
\begin{eqnarray}
    \rho_\beta=\frac{e^{-\beta \hat H}}{\sum_i e^{-\beta E_i}}\ ,
\end{eqnarray}
where $\beta$ is the inverse temperature, and the denominator is the canonical partition function.

\subsection{Optimization algorithm}

To variationally prepare the thermal state, we will employ a pulse-level version of the variational quantum thermalizer (VQT) algorithm proposed in Ref.~\cite{Selisko:2022wlc}. To the best of our knowledge, this is the first application of QOC techniques to thermal state preparation. The algorithm splits the learning process into two variational circuits where the first circuit (VQC$_1$) is dedicated to learning the probability distribution of each pure state within the ensemble, and the second circuit (VQC$_2$) learns the energy of the ensemble. With this information, the free energy $F=E-S/\beta$ is calculated, and the variational parameters are optimized to minimize $F$. Here $S$ is the entropy of the system, and $E$ is the energy. This allows us to sample from the equilibrium state at a fixed inverse temperature $\beta$. Example applications of the gate-based version of this algorithm can be found in Refs.~\cite{Araz:2024xkw, Araz:2023ngh, Fromm:2023npm}. 

We denote the unitary operators corresponding to VQC$_{1,2}$ by $\hat{U}_{1,2}(\Theta_{1,2})$ where $\Theta_{1,2}$ are the two sets of variational parameters. For a given set of variational parameters, the entropy of the system is estimated using the Shannon entropy
\begin{eqnarray}
    S = -\sum_i p_i \log p_i\ , \label{eq:shannon}
\end{eqnarray}
which is an approximation of the von Neumann entropy. Here, $p_i$ represents the probability of each pure state $\ket{\phi_i}$ estimated by VQC$_1$, $\lVert\langle \phi_i | \hat{U}_1(\Theta_1) | 0 \rangle \rVert^2$. Once the probability distribution is obtained from VQC$_1$, we compute the total energy of the ensemble as
\begin{eqnarray}
    E = \sum_i p_i \langle \phi_i | \hat{U}^\dagger_2(\Theta_2)\, \hat{H}\, \hat{U}_2(\Theta_2) | \phi_i \rangle\ , \label{eq:energy}
\end{eqnarray}
where $\{\mathbf{\Omega}_{1,2}, \mathbf{v}_{1,2}\} \in \Theta_{1,2}$. Fig.~\ref{fig:pulse_vqt_scheme} shows a schematic representation of the algorithm. Notice that compared to the algorithm in Ref.~\cite{Selisko:2022wlc}, the gates have been replaced by QOC components, where different colored pulses are used for VQC$_1$ and VQC$_2$.

\subsection{Numerical results}

\begin{figure}
    \centering
    \includegraphics[width=0.9\linewidth]{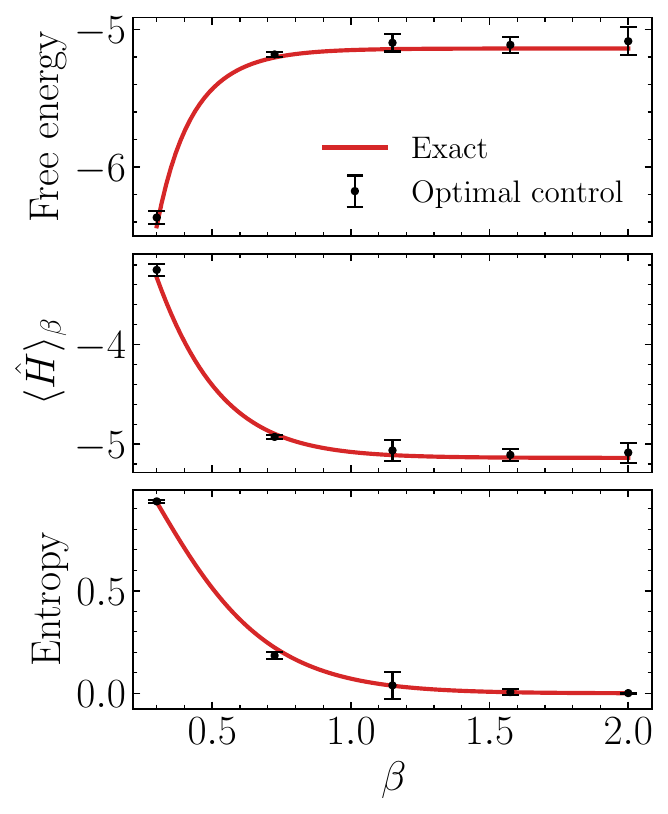}
    \caption{\it Results for the thermal state preparation of the two-site Schwinger Hamiltonian as a function of the inverse temperature $\beta$. In each panel, the red line represents the results obtained from exact diagonalization, while the black markers represent the reconstructed results. The panels display the free energy, expected energy, and entropy distributions from top to bottom. The pulse duration for this example is set to 50 ns for both variational ansätze. The uncertainty associated with the black markers indicates one standard deviation obtained from 20 random initializations of the optimization algorithm.}
    \label{fig:vqt}
\end{figure}

For simplicity, we fix the pulse duration of each circuit to 50 ns but with independent sets of parameters. We employ the same pulse setup as discussed in section~\ref{sec:state-prep-qoc}. That is, each qubit pulse consists of 100 independent segments and one phase difference that are optimized within the same control constraints. We study the ability of this algorithm to minimize the free energy of the two-site Schwinger model, where we estimate the optimization error by executing the optimization sequence for $20$ independent initializations. Fig.~\ref{fig:vqt} shows the results of the optimization process (black markers) compared to results using exact diagonalization (red line). The plot is divided into three panels, where the top panel shows the free energy as a function of $\beta$, which is the cost function for the optimization. The middle panel shows the energy estimation of the ensemble, and the lower panel shows the entropy estimation. We find the uncertainties to be rather small, providing a reliable and flexible optimization platform, which motivates more detailed investigations in future work.

To minimize the number of hyperparameters, we use a pulse duration of 50 ns for both VQC$_1$ and VQC$_2$. However, it is possible to choose a significantly shorter pulse duration VQC$_1$. The reason for choosing the same for both is due to the optimization landscape. While estimating the ground states of the same Hamiltonian, we found the shortest pulse duration to be around $30$ ns for the initial state $|10\rangle$. However, this increases to a maximum of $50$ ns for the $|01\rangle$ initial state. Hence, to enable an ansatz that is flexible enough, we chose a pulse duration of $50$ ns for the ensemble estimation.

\section{Conclusions \& Outlook~\label{sec:conclusions}}

In this work, we performed exploratory studies using Quantum Optimal Control (QOC) techniques in the context of lattice field theories relevant to fundamental physics applications. One of the critical challenges of quantum simulations of the real-time dynamics of scattering amplitudes or the non-equilibrium dynamics of finite-temperature systems is the preparation of states. Here, we explored the variational preparation of both ground and thermal states using a QOC ansatz where the parameters of a transmon control Hamiltonian are optimized at the pulse level. As a representative example, we considered the Schwinger model, which corresponds to quantum electrodynamics in 1+1 dimensions. In the Schwinger model formulation used throughout this work, the U(1) gauge field is replaced by an asymmetric long-range interaction between lattice sites, and the entire Hamiltonian can be written as a spin model. 

Using representative parameters for transmon Hamiltonians, we explored the ground state preparation of the Schwinger model for up to 4 lattice sites. While the superposition of the ground state and the long-range interaction of the Schwinger model generally pose challenges for the optimization, we found good agreement with results from exact diagonalization. We compared the time spent on state preparation using the QOC ansatz to gate-based methods. In particular, we compared a single layer of a Hamiltonian-based ansatz and an ansatz based on strongly entangling layers. While a quantitative one-to-one comparison between pulse and gate-based approaches is challenging, we observed significant speedups. We also observed that the achievable speedup obtained with our current algorithm can depend significantly on the number of qubits and device parameters. We found that, in particular, the inter-qubit coupling strength and the connectivity of the device play an important role. Aside from reduced coherence time requirements, we found that the QOC-based techniques allow for a favorable landscape of the loss function. In our simulations, we included the leakage to higher states of the transmon Hilbert space, and we studied the resilience of the optimization algorithm in the presence of device noise. The noise model is based on a time-dependent Lindblad equation that describes the evolution of the transmon qubits as an open quantum system.

Additionally, we extended our studies to the thermal state preparation of the Schwinger model. The equilibrium states of quantum systems at finite temperature values are described by a density matrix, making the corresponding quantum simulations more challenging. Starting from a quantum algorithm with intermediate measurements where the free energy is minimized, we replaced the variational ansatz based on gates with a pulse-level ansatz. For the relatively small-scale exploratory studies performed here, we obtained good agreement with results using exact diagonalization. Our results motivate further dedicated studies in this direction.

Given the near-term prospects of the noisy quantum hardware, our results indicate that QOC techniques may significantly advance quantum simulations of lattice field theories relevant to fundamental physics. Using the fact that the Schwinger model exhibits confinement, recent studies introduced scalable gate-based circuits for the variational preparation of ground states and wave packets relevant to simulations of scattering processes. These techniques may also provide a path forward to achieve a scalable algorithm for state preparation using QOC techniques, which will be the focus of future work. 

\section*{Acknowledgement}

JYA acknowledges the hospitality of IQuS at the University of Washington and the fruitful discussions during the Pulses, Qudits, and Quantum Simulations workshop. JYA and FR are supported by the U.S. Department of Energy, Office of Science, Office of Nuclear Physics with contract No.~DE-AC05-06OR23177, under which Jefferson Science Associates, LLC operates Jefferson Lab and in part by the DOE, Office of Science, Office of Nuclear Physics, Early Career Program under contract No. DE-SC0024358. TJM was supported by the U.S. National Science Foundation Research Experience for Undergraduates at Old Dominion University Grant No. 1950141.

\bibliographystyle{utphys}
\bibliography{bibliography}

\end{document}